\documentclass[3p,times,twocolumn]{elsarticle}

\usepackage{ecrc}
\newcommand{\carbon}{\rm ^{12}C}
\newcommand{\deuteron}{\rm ^{2}H}

\usepackage{array, graphicx, subfigure}
\usepackage{graphics}
\volume{00}

\firstpage{1}

\journalname{Nuclear Physics B Proceedings Supplement}

\runauth{A. Bodek et al. }

\jid{nuphbp}

\jnltitlelogo{Nuclear Physics B Proceedings Supplement}

\usepackage{amssymb}

\usepackage[figuresright]{rotating}

\begin{document}

\begin{frontmatter}



\dochead{}

\title {Effective Spectral Function for Quasielastic Scattering on Nuclei from Deuterium to Lead}

\author[label1]{A. Bodek}
\author[label2]{M. E. Christy}
\author[label1]{B. Coopersmith}

\address[label1] {Department of Physics and Astronomy, University of Rochester, Rochester, NY. 14627, USA}
\address[label2] {Department of Physics, Hampton University, Hampton, Virginia, 23668 USA  
\\Presented  by A. Bodek at  the 37th International Conference on High Energy Physics (ICHEP 2014, 2-9 Jul 2014. Valencia, Spain, [C14-07-02] )}

\begin{abstract}
Spectral functions do not fully describe quasielastic electron and neutrino scattering from nuclei because they only model the initial state.  Final state interactions distort the shape of the differential cross section at the peak and  increase the cross section  at the tails of the distribution.  We show that  the  kinematic distributions predicted by  the $\psi'$ superscaling formalism  can be well described  with a modified    {\it {effective spectral function}} (ESF).   By construction, models using ESF in combination with  the transverse enhancement contribution correctly predict  electron QE  scattering data. Our values for the binding energy parameter $\Delta$  are smaller than  $\overline{\epsilon}$ extracted within the Fermi gas model from  pre 1971 data by Moniz\cite{moniz}, probably  because these early cross sections  were  not corrected for coulomb effects.

\end{abstract}


\end{frontmatter}



%
\section{Introduction}
\label{intro}
Neutrino oscillation experiments make use of neutrino Monte Carlo (MC) event generators to model the  cross sections and  kinematic distributions of the leptonic and hadronic final state of neutrino interactions on nuclear targets.   Because of the conservation of the vector current (CVC), the same models should be able to  reliably predict the  quasielaststic  electron scattering cross section on nuclear targets. Unfortunately, none of  the models that are currently implemented in neutrino MC generators are able to do it.  Here we summarize an approach which guarantees agreement with  QE  electron  scattering data by construction. A more detailed
description is given in reference \cite{epic-paper}

\begin{figure}[ht]
\begin{center}
\includegraphics[width=1.5in,height=1.4in]{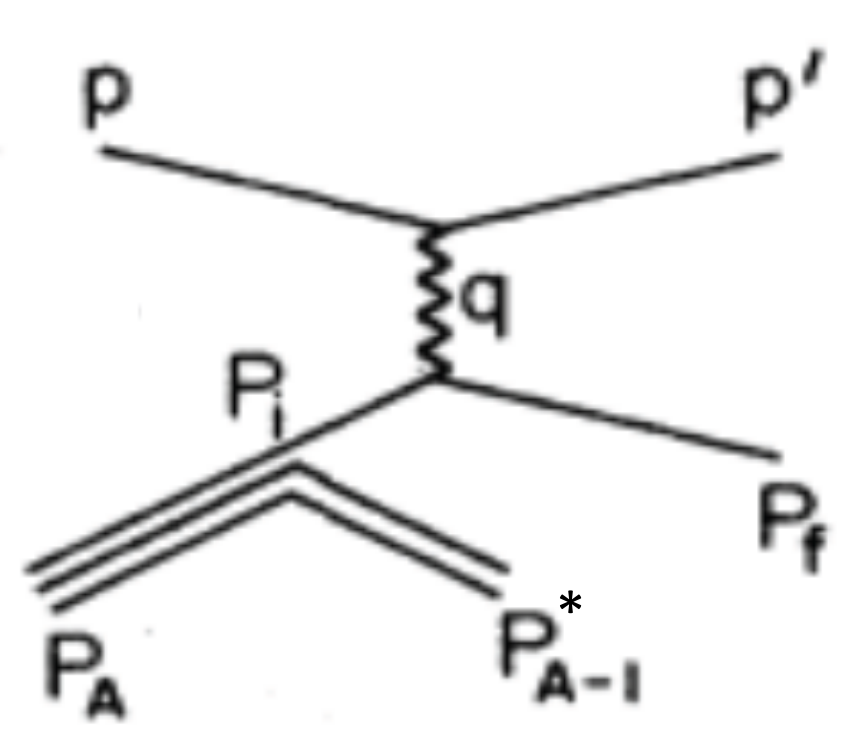}
\includegraphics[width=1.5in,height=1.4in]{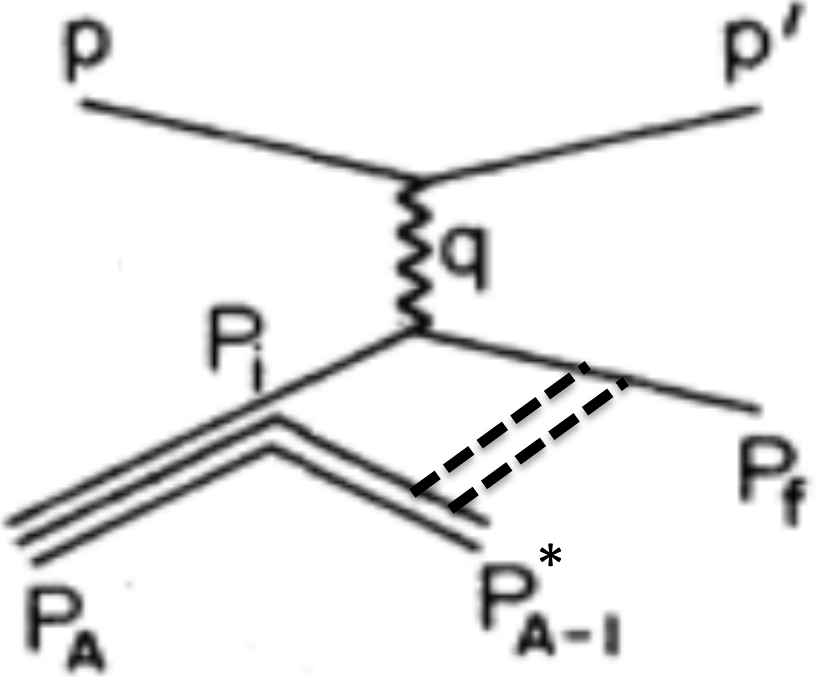}
\caption{ Left: Scattering from an off-shell bound neutron of momentum $\bf{P_i}=k$ in 
a nucleus of mass A.  The on-shell recoil  $[A-1]^*$ (spectator) nucleus has a momentum $\bf{P_{A-1}^*= P_s=-k}$. This process
is referred to as the 1p1h process (one proton one hole).
Right:  The 1p1h process including final state interaction (of the first kind) with another nucleon.  
}
\label{Aoff-shell}
\end{center}
\end{figure}

The  left panel of Fig. \ref{Aoff-shell} is the general diagram for QE lepton (election, muon or neutrino)  scattering from a nucleon which is bound in a nucleus of mass $M_A$.  In this paper, we focus on charged current neutrino scattering.
The scattering is from an off-shell bound neutron of momentum $\bf{P_i}=k$.  The on-shell recoil  $[A-1]^*$ (spectator) nucleus has a momentum $\bf{P_{A-1}^*= P_s=-k}$. This process
is referred to as the 1p1h process (one proton one hole).
The * is used to indicate that the spectator nucleus is not in the ground state because it has one hole.
The four-momentum transfer to the nuclear target is defined as $q = (\vec q,  \nu)$.  Here  $\nu$ is the energy transfer, and
$Q^2= -q^2 = \nu^2- \vec{q}^2$ is the square of the four-momentum transfer.  For free nucleons the energy transfer $\nu$ is
equal to $Q^2/2M_N$ where $M_N$ is the mass of the nucleon. At a fixed value of $Q^2$,  QE scattering  on nucleons bound in a nucleus yields a distribution in $\nu$ which peaks at  $\nu=Q^2/2M_N$.  In this communication, the term
"normalized quasielastic distribution" refers to the normalized  differential cross section $\frac{1}{\sigma} \frac{d\sigma}{d\nu}(Q^2,\nu)=\frac{d^2\sigma /dQ^2 d\nu} {<d\sigma/dQ^2>} $ where $<\frac{d\sigma}{d Q^2}>$ is  the integral of $[\frac{d^2\sigma}{dQ^2 d\nu } ]d\nu$ over all values of $\nu$ (for a given value of $Q^2$).

%

\begin{figure}
\begin{center}
\includegraphics[width=2.5in,height=2.2in]{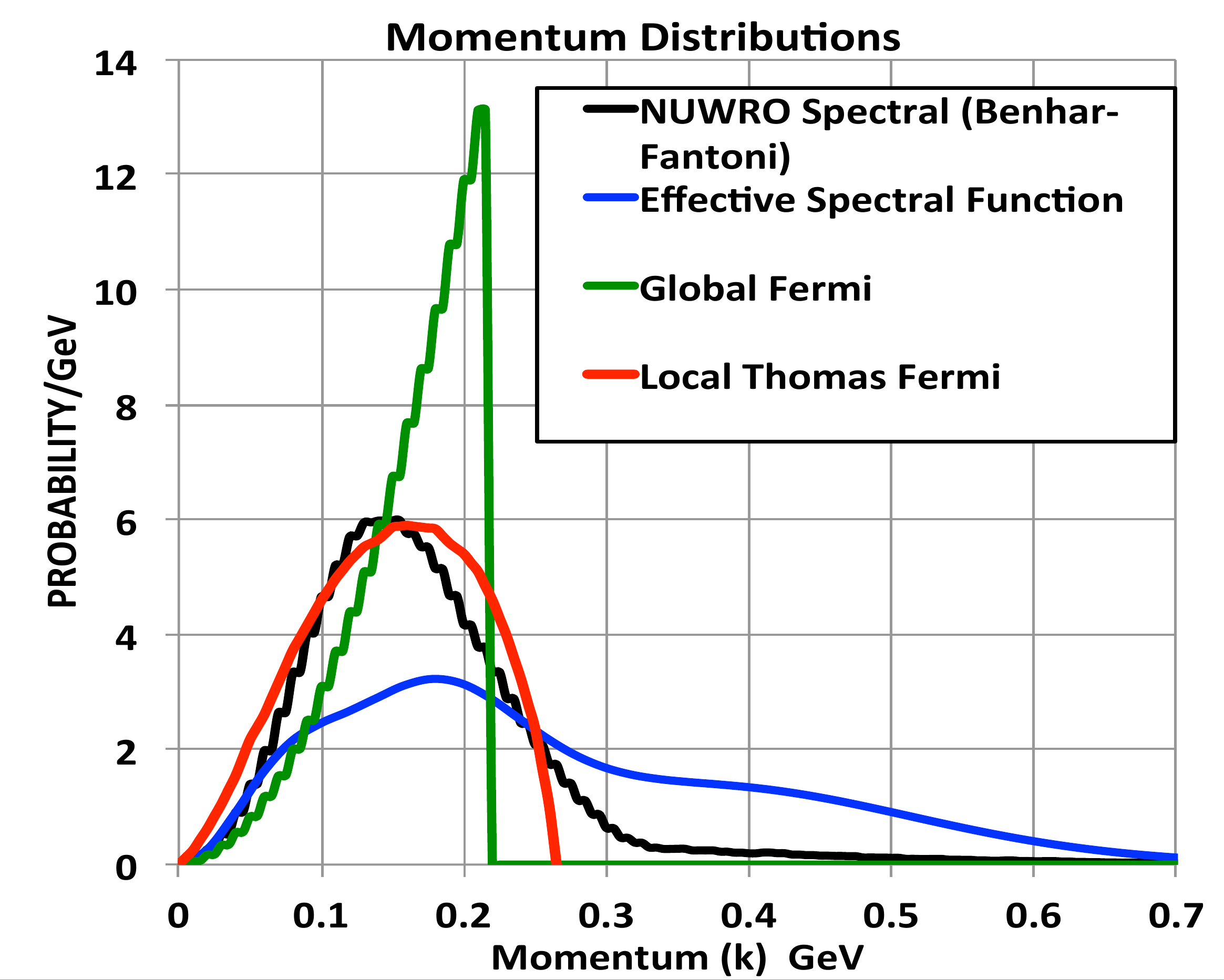}
\caption{ 
Nucleon momentum distributions  in a $\carbon$ nucleus for several spectral functions. The curve labeled "Global Fermi" gas is
the momentum distribution for the Fermi gas model.
The blue line is the momentum distribution for the  {\it {effective spectral function}} described in this paper.
}
\label{momentum}
\end{center}
\end{figure}
The  right  panel of Fig. \ref{Aoff-shell} shows the same QE lepton scattering process, but now also  including  a final state interaction  with another nucleon in the scattering process. This final state interaction modifies
the scattering amplitude and  therefore can change the kinematics of the final state lepton. In this paper, we refer to it as 
"final state interaction of the first kind" (FSI).
The final state nucleon can then undergo more  interactions with other nucleons
in the spectator nucleus.  These  interactions do not change the energy of the final state lepton.  We refer to these
 final state interactions as "final state interaction of the second kind".


 In general, neutrino event generators assume that the  scattering occurs on  independent nucleons which are bound in the nucleus.
Generators such as  GENIE\cite{genie}, NEUGEN\cite{neugen}, NEUT\cite{neut}, NUANCE\cite{nuance} NuWro \cite{nuwro} and GiBUU\cite{gibuu}   account for nucleon binding effects  by modeling the momentum distributions and removal energy of nucleons in nuclear targets. Functions that describe the momentum distributions and removal energy of nucleons from nuclei are referred to as spectral functions. Spectral functions describe the initial state.

 Spectral functions can take the  simple form of a momentum distribution and a fixed removal energy (e.g. Fermi gas\cite{moniz}),  or the more complicated form of a two dimensional (2D)  distribution in  momentum and removal energy (e.g. Benhar-Fantoni spectral function \cite{bf}).  

Fig. \ref{momentum} shows  the nucleon momentum distributions in a $\carbon$ nucleus for some of the
spectral functions that are currently being used. The solid green line is the nucleon  momentum distribution for the  
 Fermi gas\cite{moniz}  which is currently implemented in all neutrino event generators. The solid black line is  the projected momentum distribution of the  Benhar-Fantoni \cite{bf}  2D spectral function as implemented in NuWro.  
The solid red line is the nucleon momentum distribution for  the Local-Thomas-Fermi gas (LTF).

It is known that  theoretical calculations using spectral functions do not fully describe the shape of the  quasielastic  peak for  electron scattering on nuclear targets .  This is  because the calculations  only model the initial state (shown on the left panel of Fig. \ref{Aoff-shell}), and do not account for final state interactions of the first kind  (shown on the right panel of Fig. \ref{Aoff-shell}) .
Because FSI changes the amplitude of the scattering,  it modifies the shape of $\frac{1}{\sigma}\frac{d\sigma}{d\nu}$. FSI reduces the cross section at the peak and  increases the cross section  at the tails of the distribution.
\begin{figure}[ht]
\begin{center}
\includegraphics[width=2.99in,height=3.in]{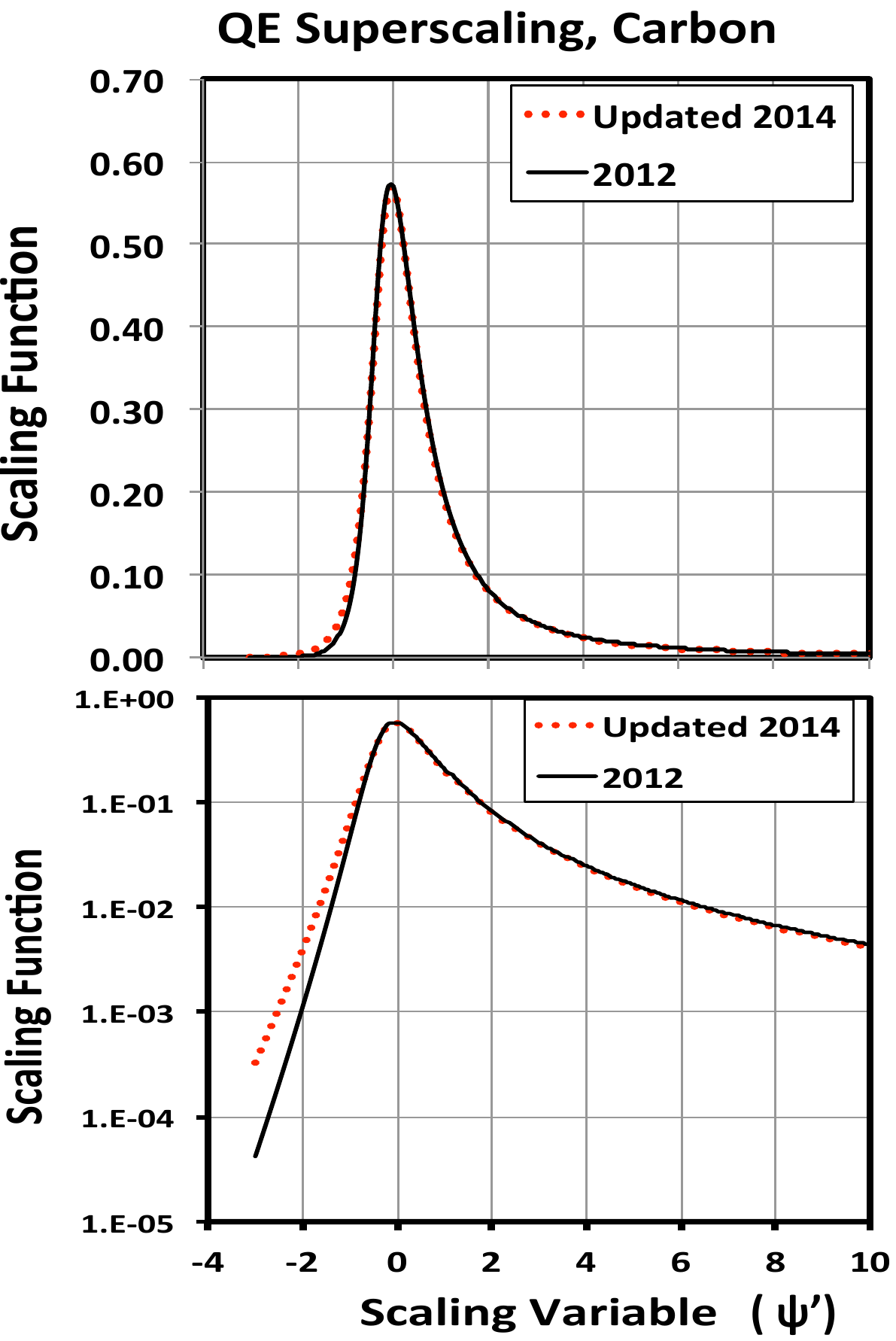}
\caption{  The  $\psi'$ superscaling distribution extracted from a fit to electron scattering data used by  Bosted and Mamyan \cite{super2} (solid black) labeled as 2012, and the  superscaling function  extracted from a more recent updated fit \cite{eric}  to data from a large number of quasielastic  electron scattering experiments  on $\carbon$ (dotted red) labeled as 2014). The top panel shows the superscaling functions on a  a linear scale. The bottom panel  shows the same superscaling functions on  a logarithmic scale. The integral of the curve has been normalized to unity.
}
\label{superfunction}
\end{center}
\end{figure} 
In contrast to the spectral function formalism, predictions using the  $\psi'$ superscaling formalism\cite{super1,super2} fully describe the longitudinal response function of  quasielastic electron scattering data on nuclear targets.  This is expected since the calculations use a   $\psi'$ superscaling function  which is  directly extracted from the  longitudinal  component of measured electron scattering QE differential cross sections.
 
In this communication  we present the parameters for a new {\it {effective spectral function}}  that reproduces the  
 kinematics of the final state lepton  predicted by $\psi'$ superscaling.  The momentum distribution for this ESF for  $\carbon$ is shown as the  blue line in Fig. \ref{momentum}.


 %
\subsection { The $\psi'$ superscaling functions for QE scattering}
The  $\psi$ scaling variable\cite{super1,super2} is defined as:
\begin{equation}
\psi\equiv \frac{1}{\sqrt{\xi_F}} \frac{\lambda-\tau}{\sqrt{(1+\lambda)\tau+
\kappa\sqrt{\tau(1+\tau)}}},
\label{eq:psi}
\end{equation} 
where  $\xi_F\equiv \sqrt{1+\eta_F^2}-1$,  $\eta_F \equiv K_F/M_n$, $\lambda \equiv\nu /2M_n$, $\kappa \equiv {|\vec q| }/2M_n$ and $\tau \equiv|Q^{2}|/4M_n^{2}=\kappa ^{2}-\lambda ^{2}$. 

The  $\psi'$ superscaling variable includes
a correction that accounts for the removal energy from the nucleus. This is achieved by
replacing  $\nu$ with $\nu-E_{\mathrm{shift}}$, which  forces the maximum of the QE response to occur at 
$\psi^\prime=0$.
QE scattering on all nuclei (except for the deuteron) is described using the same universal superscaling function. The only
parameters which are specific to each nucleus are   the Fermi broadening parameter $K_F$ and the  
 energy  shift  parameter  $E_{\mathrm{shift}}$.

  Fig. \ref{superfunction} shows  two parametrizations of  $\psi'$ superscaling functions extracted from quasielastic  electron scattering data  on $\carbon$.  Shown  is the  $\psi'$ superscaling distribution extracted from a fit to electron scattering data used by  Bosted and Mamyan \cite{super2} (solid black line labeled as 2012), and the  superscaling function  extracted from a recent updated fit\cite{eric}  to data from a large number of quasielastic  electron scattering experiments  on $\carbon$ (dotted red line labeled as 2014). The  top panel  shows the superscaling functions on a  a linear scale and the bottom panel shows the same superscaling functions on  a logarithmic scale.   
  
The  $\psi'$ superscaling function is extracted from the longitudinal QE cross section
for $Q^2>0.3$~GeV$^2$ where there are no Pauli blocking effects. At very 
low values of $Q^2$, the  QE differential
cross sections predicted by the   $\psi'$ superscaling should be multiplied by
a Pauli blocking factor $K_{Pauli}^{nuclei}(Q^2)$
which   reduces the predicted cross sections at low $Q^2$. The Pauli suppression
factor is  given\cite{super2} by the function
\begin{equation}
K_{Pauli}^{nuclei} =\frac{3}{4} \frac{|\vec q|}{K_F}(1 - \frac{1}{12} (\frac {|\vec q|}{K_F})^2)
\label{nucpauli}
\end{equation}
For $ |\vec q| < 2K_F$, otherwise no Pauli suppression correction is made.  Here  $ |\vec q| =\sqrt{Q^2+\nu^2}$ is the absolute magnitude of the momentum transfer to the target nucleus,

 

 %

\begin{figure}
\includegraphics[width=2.99in,height=2.5in]{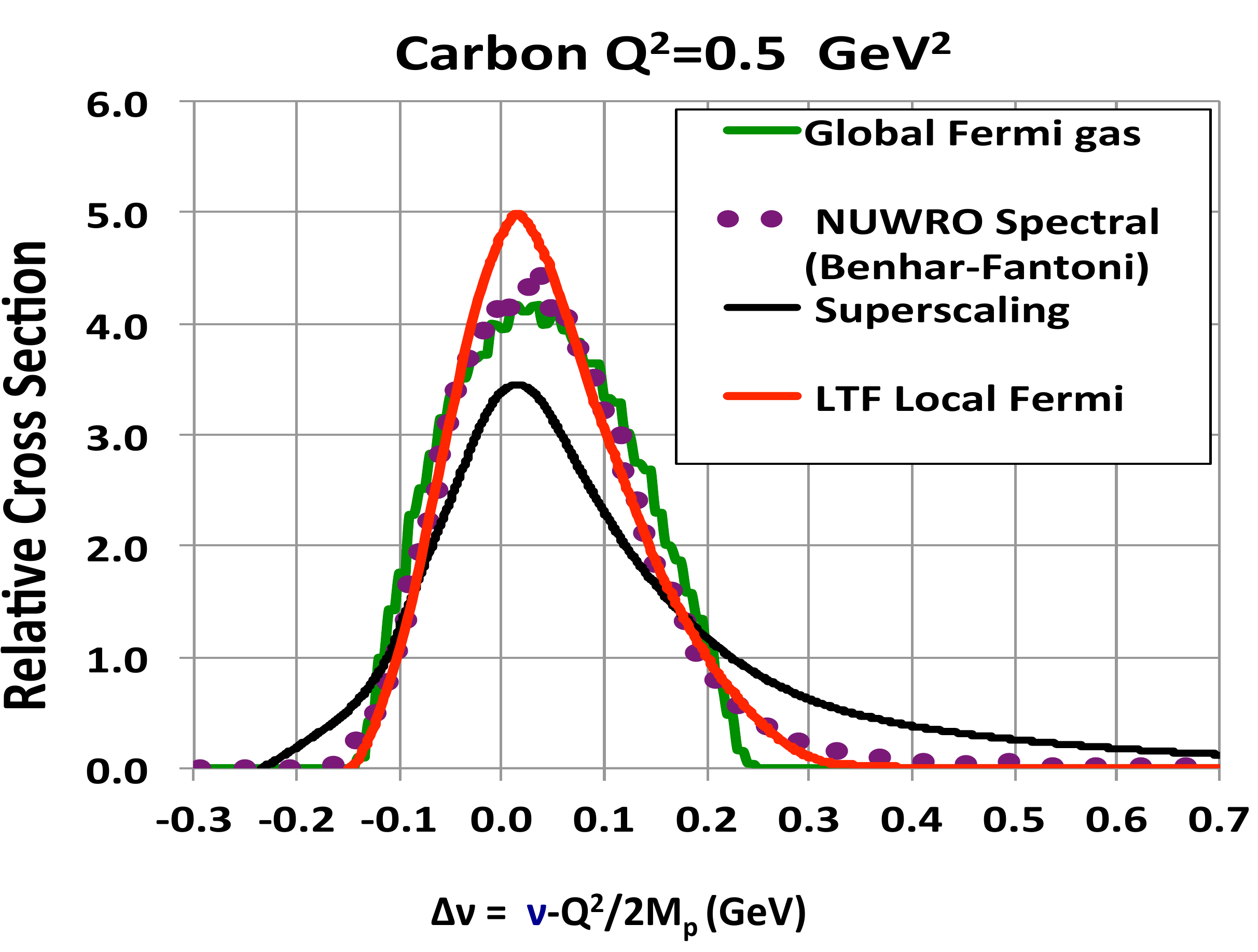}
\caption{ Comparison of the  $\psi'$ superscaling prediction (black line)  for  the normalized  $\frac{1}{\sigma} \frac{d\sigma}{d\nu}(Q^2,\nu)$  at $Q^2$=0.5 GeV$^2$ for 10 GeV neutrinos on $\carbon$ 
to the predictions of several spectral function models.  Here   $\frac{1}{\sigma} \frac{d\sigma}{d\nu}(Q^2,\nu)$ 
is plotted versus  $\Delta \nu$
 The  
  predictions of  the  spectral function models are in disagreement with the predictions
of  $\psi'$ superscaling.
.
}
\label{nuwro-vs-scaling}
\end{figure} 
%
\begin{table}
\centering
\begin{tabular}{ccccccccc}
\hline
$A$ & $K_F(\psi')$ (GeV) & $E_{\mathrm{shift}}(\psi')$ (GeV) \\ \hline
2 & 0.100 & 0.001 \\
3 & 0.115 & 0.001 \\
$3<A<8$ & 0.190 & 0.017 \\
$7<A<17$ & 0.228 & 0.0165 \\
$16<A<26$ & 0.230 & 0.023 \\
$25<A<39$ & 0.236 & 0.018 \\
$38<A<56$ & 0.241 & 0.028 \\
$55<A<61$ & 0.241 & 0.023 \\
$A>60$ & 0.245 & 0.018 \\
\hline
\hline
\end{tabular}
\caption{Values of Fermi-broadening parameter $K_F$
and  energy shift $E_{\mathrm{shift}}$ used in the   $\psi'$ superscaling prediction for different nuclei.
The parameters for deuterium (A=2) are  a crude approximation only, and
deuterium is treated differently as discussed in reference \cite{epic-paper}.}
\label{tab:kfes}
\end{table}
%
\subsection{Comparison of models for quasielastic scattering}

Fig. \ref{nuwro-vs-scaling}  
shows predictions for the normalized  QE  differential cross sections 
   $\frac{1}{\sigma} \frac{d\sigma}{d\nu}(Q^2,\nu)$ for 10 GeV neutrinos on $\carbon$ at  $Q^2$=0.5 GeV$^2$ for various
   spectral functions.    Here  $\frac{1}{\sigma}\frac{d\sigma}{d\nu}$ 
is plotted versus  $\Delta\nu=\nu-\frac{Q^2}{2M_p}$. 
The prediction of the  $\psi'$ superscaling formalism for   $\frac{1}{\sigma} \frac{d\sigma}{d\nu}(Q^2,\nu)$ is shown  as the solid black line. 
  The solid green line is the prediction using the  "Global Fermi" gas \cite{moniz}.
 The solid red  line  is the prediction using
the Local  Thomas Fermi gas (LTF)  momentum distribution. 
The dotted purple line is the NuWro prediction using the full two dimensional Benhar-Fantoni\cite{bf}
spectral function.  
 The predictions of all of these spectral functions  for   $\frac{1}{\sigma} \frac{d\sigma}{d\nu}(Q^2,\nu)$  are in disagreement with the predictions
of  the  $\psi'$ superscaling formalism.
%
%

%
\section{Effective Spectral Function for $\carbon$}
%
\subsection{Momentum Distribution}
The probability distribution for a nucleon to have
a momentum $k=|\vec{k}|$ in the nucleus is defined as  
$$P(k) dk=4\pi k^2 |\phi(k)|^2dk.$$
For $k<0.65$ GeV, we  parametrize\cite{bfit} $P(k)$  by the following function:
\begin{eqnarray}
P(k) &=& \frac{\pi}{4c_0}\frac{1}{N}(a_s+a_p+a_t)y^2 \\
y &=&\frac{k}{c_0}; ~~~~~~a_s =c_1  e^{-(b_sy)^2 } \nonumber\\
a_p &=& c_2(b_py)^2 e^{-(b_py)^2};  ~~~a_t= c_3 y^\beta e^{-\alpha(y-2)} \nonumber
\end{eqnarray}
For $k>0.65$ GeV we set $P(k)$ = 0. 
Here,  $c_0=0.197$,  $k$ is in GeV, N is a normalization factor to normalize
the integral of the momentum distribution  from $k$=0 to $k$=0.65 GeV to 1.0,  and P($k$) is in units of  GeV$^{-1}$.
%

\begin{figure}
\begin{center}
\includegraphics[width=1.4in,height=1.4in]{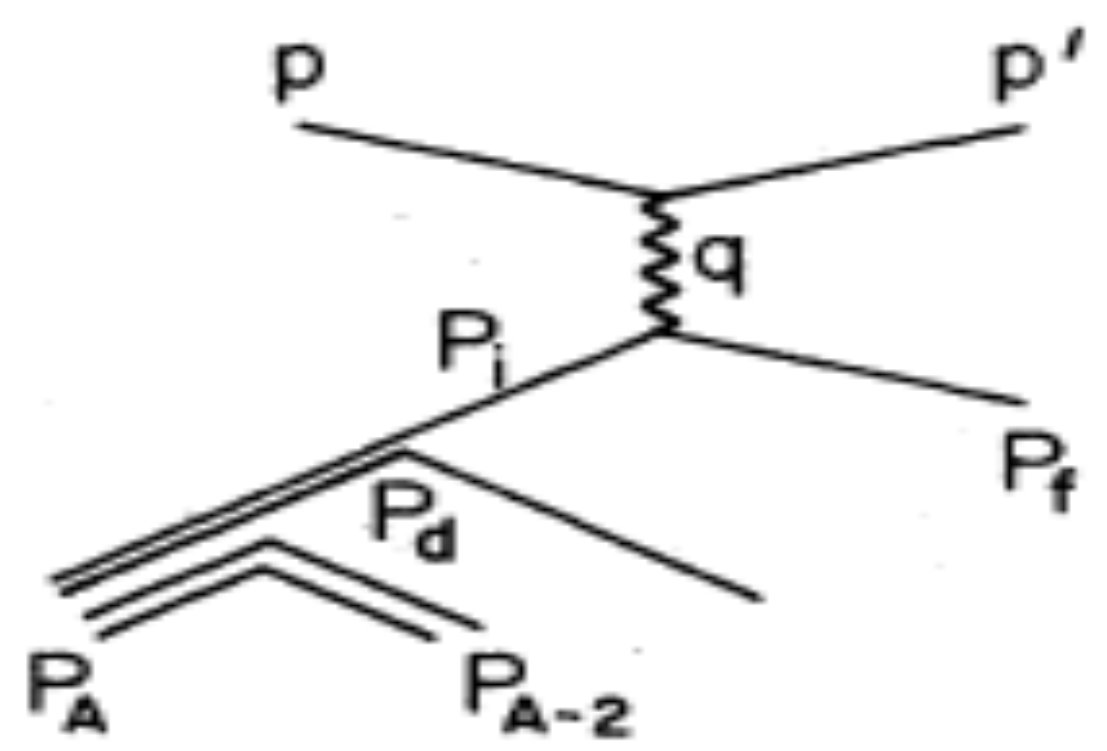}
\caption{ 2p2h process:  Scattering from an off-shell bound neutron of momentum $\bf{P_i=-k}$ from two nucleon
correlations (quasi-deuteron).   The on-shell recoil  spectator nucleon has momentum  $\bf{P_s=k}$.}
\label{Doff-shell}
\end{center}
\end{figure}
\subsection{Removal Energy}
The kinematics for neutrino charged current quasielastic 
 scattering from a off-shell bound neutron with momentum
$\bf {k}$ and energy  $E_n$ are given by:
\begin{eqnarray}
\label{eq-nu}
(M_{n}')^2 &= & (E_n)^2 - {Vk^2} \\
M_p^2 &= &(M_{n}')^2 + 2E_n\nu-2|\vec q| k_z- Q^2  \nonumber\\
\nu &= & E_\nu - E_\mu  =   \frac{Q^2+M_p^2- (M_{n}')^2 + 2|\vec q |k_z}{(E_n)} \nonumber\\
V(Q^2) &= &  1 -e^{-xQ^2},~~x=12.04
\label{Veq}
\end{eqnarray}

For scattering from a single off-shell nucleon, the term $V(Q^2)$ multiplying $k^2$  in Equations \ref{eq-nu}, \ref{En1p1h},  and  \ref{En2p2h} should be 1.0.  However, we find that in order to make the spectral function predictions agree with $\psi'$ superscaling at very low $Q^2$ (e.g. $Q^2<0.3~GeV^2$) we need to apply a $Q^2$-dependent correction  to reduce the removal energy, e.g. due to  final state interaction (of the first kind) at low $Q^2$. This factor is given in equation \ref{Veq}. 
 The value of the parameter $x$=12.04 GeV$^{-2}$  was extracted from the fits at low
 values of  $Q^2$. As mentioned earlier, $\vec q$ is the  momentum transfer to the neutron. 
We define the component of the initial neutron momentum $\bf k$ which is
parallel to $\vec q$ as $k_{z}$.
The expression for   $E_n$ depends  on the process and is given by  Equations \ref{En1p1h} and  \ref{En2p2h} for the 1p1h, and 2p2h
process, respectively.


We  assume
 that the off-shell energy  ($E_n$)  for a  bound neutron with momentum
 $\bf{k}$ can  only take two possible values\cite{Bodek-Ritchie}. We refer to the 
 first possibility  as the 1p1h process (one proton, one hole  in the final state).
 The second possibility is the 2p2h  process(two protons and two holes in the final state).
 
    In our {\it{effective spectral function}} model the 1p1h process occurs
 with probability $f_{1p1h}$, and the 2p2h process   occurs with
 probability of $1-f_{1p1h}$.  For simplicity, we assume that  the probability  $f_{1p1h}$
 is independent of  the momentum 
 of the bound nucleon.
 \subsubsection{The 1p1h process}
The 1p1h process refers to scattering from an independent neutron in the nucleus resulting in a final state proton and
a hole in the spectator nucleus.
Fig. \ref{Aoff-shell} illustrates the  1p1h process  (for $Q^2>0.3$ GeV$^2$),
for the  scattering from an off-shell bound neutron of momentum $\bf{-k}$ in a nucleus of mass A\cite{Bodek-Ritchie}.   In the 1p1h process, momentum is balanced by an  on-shell recoil  $[A-1]^*$  nucleus  which has momentum $\bf{P_{A-1}^*}=\bf{P_s}=k$ and an average binding energy 
parameter $\Delta$, where  $ M_{A}-M_{A-1}^*=M_n +\Delta$.  The initial state off-shell neutron has energy $E_n$ which is given by:
\begin{eqnarray}
E_n (1p1h)&= & M_A - \sqrt{Vk^2+(M_{A-1}^*)^2} \nonumber \\
&\approx& M_n-\Delta-\frac{Vk^2}{2M_{A-1}^*}
\label{En1p1h}
\end{eqnarray}
The final state includes a proton and an $[A-1]^*$ nucleus in an excited state because the removal of the nucleon
leaves  hole in the energy levels of the nucleus.

 \begin{figure}
\begin{center}
\includegraphics[width=2.99in,height=2.1in]{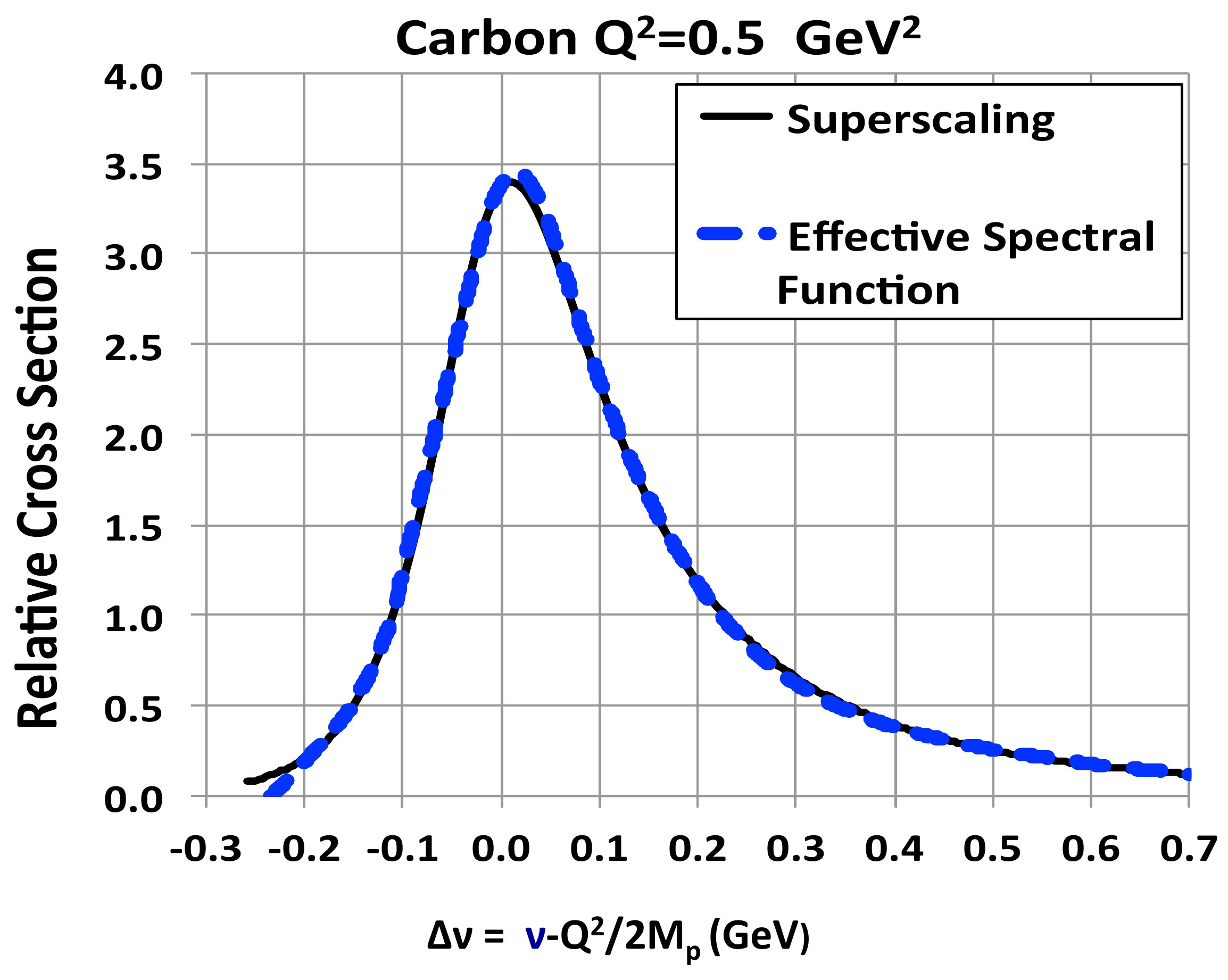}
\caption{ Comparison of the prediction for the normalized QE differential cross section   ($\frac{1}{\sigma} \frac{d\sigma}{d\nu}(Q^2,\nu)$)   for $\carbon$
from the  {\it{effective spectral function}}
to the prediction of   $\psi'$ superscaling. The predictions 
are shown as
a function of $\Delta \nu$ at  $Q^2$=0.5 GeV$^2$. 
For $Q^2$=0.5 GeV$^2$  the prediction
of the  {\it{effective spectral function}} are almost identical to the prediction of  $\psi'$  superscaling. 
}
\label{finaleffective}
\end{center}
\end{figure} 
%


  \subsubsection{ The 2p2h process}
In general, there are several processes which result in two  (or more) nucleons and a spectator  excited nucleus with two (or more)  holes in final state:
\begin{itemize}
\item  Two nucleon correlations in initial state (quasi deuteron) which are  often  referred to as short range correlations (SRC).
\item  Final state interaction  (of the first kind) resulting in a larger energy transfer to the hadronic final state (as modeled by superscaling).
\item  Enhancement of the transverse cross sections ("Transverse Enhancement") from meson exchange currents
(MEC) and isobar excitation.
\end{itemize}

In the {\it{effective spectral function}}  approach the lepton energy spectrum  for all three processes is modeled as originating from the  two nucleon correlation process. This accounts for the additional energy shift resulting from the removal of two nucleons from the nucleus.

 Fig. \ref{Doff-shell} illustrates the  2p2h process for  scattering from an off-shell bound neutron of momentum $\bf{-k}$ (for $Q^2>0.3$ GeV$^2$).
The momentum of the interacting nucleon in the initial state
 is balanced by a single on-shell correlated recoil  nucleon  which has momentum $\bf{k}$. The $[A-2]^*$ spectator nucleus is left
 with two holes. The initial state off-shell neutron has energy $E_n$ given by:
\begin{eqnarray}
E_n (2p2h)&= & (M_p+M_n) - 2\Delta -\sqrt{Vk^2+M_{p}^2} 
\label{En2p2h}
\end{eqnarray}
\noindent where  $V$  is given by eq. \ref{Veq}.

\begin{table}
\begin{center}
\begin{tabular}{|c|c|c|c|} \hline 
 &  Benhar-    &  ESF &  ESF \\
  &  Fantoni    &  ESF &  ESF \\
 \hline\hline
Nucl.&  $\carbon$   &  $\carbon$  &  $\deuteron$  \\ \hline
$\Delta$ (MeV)  & 2Dspectral        &  12.5   &   0.13   \\ \hline
$f_{1p1h}$  & 2Dspectral        &  0.808   &  0   \\ \hline
$f_{2p2h}$  & 2Dspectral        &  0.192  &  1.00    \\ \hline
$b_s$      &  1.7                &     2.12 &  0.413475    \\ \hline
$b_p$      &  1.77              &  0.7366  &1.75629\\ \hline
$\alpha$  &  1.5                &   12.94  & 8.29029  \\ \hline
$\beta$    &  0.8               &  10.62   & 3.621 x10$^{-3}$   \\ \hline
$c_1$       &  2.823397    &  197.0   & 0.186987\\ \hline
$c_2$       &  7.225905    &  9.94     &  6.24155\\ \hline
$c_3$       & 0.00861524 & 4.36 x10$^{-5}$       & 2.082 x10$^{-4}$\\ \hline
$N$          &  0.985          &  29.64     &  10.33  \\ \hline
\end{tabular}
\caption{ A comparison of the parameters that describe the projected momentum distribution for  the Benhar-Fantoni spectral function for  $\carbon$ (2nd column) with  the parameters that describe the {\it{effective spectral function}} (ESF) (3rd column).  Here, $\Delta$ is the average binding energy  parameter of the spectator  one-hole nucleus  for the  1p1h process and $f_{1p1h}$ is the fraction of the scattering that occurs via the 1p1h process. 
For the  2p2h process the average binding energy for the two-hole spectator nucleus  is $2\Delta$.  The parameters for the  {\it{effective spectral function}} for deuterium  ($\deuteron$) are given in the 4th column.}
\label{fitsC} 
\end{center}
\end{table}
\begin{table*}
\begin{center}
\begin{tabular}{|c|c|c|c|c|c|c|c|} \hline 
Param.     & $\rm ^{4}_2He$    &  $\rm ^{12}_{~6} C$   & $\rm ^{20.8}_{~~10}Ne$   & $\rm ^{26.98}_{~~13}Al$  &  $\rm ^{39.95}_{~~18}Ar$ &  $\rm ^{55,85}_{~~26}Fe$ &
  $\rm ^{207.2}_{~~~82}Pb$ \\ \hline\hline
$f_{1p1h}$     &  0.791   &    0.808  & 0.765    & 0.774  & 0.809    & 0.822     & 0.896  \\ \hline
$b_s$          &    2.14   &   2.12    & 1.82    & 1.73     & 1.67      & 1.79       & 1.52  \\ \hline
$b_p$          &   0.775  &  0.7366   & 0.610   & 0.621 & 0.615  & 0.597    & 0.585  \\ \hline
$\alpha$         & 9.73   &   12.94   & 6.81 & 7.20 & 8.54 & 7.10 & 11.24   \\ \hline
$\beta$           &   7.57  &  10.62        & 6.08 & 6.73  & 8.62    &  6.26     & 13.33 \\ \hline
$c_1$             &  183.4 &  197.0       & 25.9  & 21.0   & 200.0  & 18.37 & 174.4   \\ \hline
$c_2$             & 5.53  &  9.94             & 0.59 & 0.59   & 6.25 & 0.505     & 5.29  \\ \hline
$c_3$               &  59.0x10$^{-5}$ & 4.36 x10$^{-5}$      & 221. x10$^{-5}$   & 121.5 x10$^{-5}$  & 269.0x10$^{-5}$ & 141.0 x10$^{-5}$ & 9.28x10$^{-5}$   \\\hline
$N$             & 18.94   &  29.64         & 4.507  & 4.065   & 40.1  & 3.645  & 37.96     \\  \hline
$\Delta$(MeV)           &   14.0    &  12.5      & 16.6      & 12.5    & 20.6     & 15.1        & 18.8   \\ \hline  \hline
$E_{shift}~\psi'$\cite{super2}           &   17.0    &  16.5      & 23.0      & 18.0    & 28.0     & 23.0       & 18.0  \\  
$K_F ~\psi'$\cite{super2}             & 190   &  228         & 230  & 236   & 241  & 241  & 245 \\
  2012          & $3<A<8$   &  $3<A<8$         & $7<A<17$  & $25<A<39$  &$38<A<56$  & $55<A<61$  & $60<A$ \\
\hline \hline
$\overline{\epsilon}$~~~Moniz\cite{moniz}           &   17.0  &  25.0      & 32.0    & 32.0           & 28.0    & 36.0        & 44.0   \\
$K_F$  ~Moniz\cite{moniz}            & 169 &  221        & 235  & 235   & 251  & 260  & 265 \\ \
  (1971)        &  $\rm ^{6.94}_{~~3}Li$   &  $\rm ^{12}_{~6} C$  & $\rm ^{24.31}_{~~12}Mg$   & $\rm ^{24.31}_{~~12}Mg$  &  $\rm ^{40.08}_{~~20}Ca$ &  $\rm ^{58.7}_{~28}Ni$ &    $\rm ^{207.2}_{~~~82}Pb$ \\ \hline
\end{tabular}
\caption{ Parameterizations of the  {\it{effective spectral function}}  for various nuclei. 
Here, $\Delta$ is the binding energy parameter, and
$f_{1p1h}$ is the fraction of the scattering that occurs via the ${1p1h}$ process.  The parameters
for    $\rm ^{3}He$ are given in reference \cite{epic-paper}.   For deuterium ($\deuteron$) see Table \ref{fitsC}, and reference \cite{epic-paper}.
The best fit values for the binding energy parameter $\Delta$  for EFS are similar but not identical to the  $E_{shift}$ parameter in the $\psi'$ scaling formalism.\cite{super2}. The EFS values for the binding energy parameter $\Delta$  are smaller than  $\overline{\epsilon}$ extracted within the Fermi gas model from 
pre 1971 electron scattering data by Moniz\cite{moniz}, probably  because these early cross sections  were  not corrected for coulomb effects.
}
\label{fitsA}
\end{center}
\end{table*}

In the  {\it{effective spectral function}}  approach,  all effects of final state interaction (of the first kind)  are absorbed
in the initial state {\it {effective spectral function}}.
 The parameters of
 the {\it {effective spectral function}} are obtained 
 by finding the parameters  $x$, 
$\Delta$, $f_{1p1h}$,  $b_s$, $b_p$, $\alpha$,  $\beta$, $c_1$, $c_2$, $c_3$, $N$ and $f_{1p1h}$     for which the predictions of the     {\it {effective spectral function}}  best describe 
the predictions of  the $\psi'$ superscaling formalism for $(1/\sigma)d\sigma/d\nu$  at  $Q^2$ values of 0.1, 0.3, 0.5 and 0.7  GeV$^2$. 

Fig. \ref{finaleffective} compares predictions for  $\frac{1}{\sigma} \frac{d\sigma}{d\nu}(Q^2,\nu)$  for $\carbon$
as  a function of $\Delta\nu$ at $Q^2$=0.5 GeV$^2$.
The prediction of  the   {\it{effective spectral function}} is the dashed blue curve.
The prediction of the    $\psi'$ superscaling model is the solid black curve. 
For $Q^2$=0.5 GeV$^2$  the prediction
of  the  {\it{effective spectral function}} is almost identical to the prediction of  $\psi'$  superscaling.

%
 We find that the  {\it{effective spectral function}}  
with only the 1p1h process provides a reasonable description of the prediction of  $\psi'$ superscaling.  Including a  contribution from the 2p2h process in the fit  improves the agreement and results in  a prediction which is almost identical to the prediction of  $\psi'$ superscaling.

The  parameterizations of the  {\it{effective spectral function}}  for all nuclei  from deuterium to lead  are given in  Tables \ref{fitsC} and  \ref{fitsA}.
As shown in Table \ref{fitsA}, the EFS values for the binding energy parameter $\Delta$  are smaller than  $\overline{\epsilon}$ extracted within the Fermi gas model from 
pre 1971 electron scattering data by Moniz\cite{moniz}.  This may be because these early cross sections were  not corrected for coulomb effects..
%
\section{Conclusion}
We present parameters for an  {\it{effective spectral function}} that reproduce
the prediction for  
 $\frac{1}{\sigma} \frac{d\sigma}{d\nu}(Q^2,\nu)$ from the $\psi'$ formalism.
We present parameters for a large number of nuclear targets
from  deuterium to lead.

  Since most of the currently 
available neutrino MC event generators model neutrino scattering in terms
of spectral functions,  the {\it{effective spectral function}} can easily be implemented.
For example, it has taken only a few days to implement
the    {\it{effective spectral function}} as an 
option in recent private  versions of NEUT and GENIE.
The predictions  for QE scattering on nuclear targets using EFS with the inclusion of the the Trasverse Ehancement\cite{epic-paper,TE}  contribution fully describe electron scattering data by construction.

\end{document}